\documentclass[aps,prb,twocolumn,showpacs,superscriptaddress]{revtex4-1}
\usepackage{graphicx}
\usepackage[final,dvips]{epsfig}
\usepackage{color}
\usepackage{dcolumn}  
\usepackage{bm}       
\usepackage{array}    
\usepackage{amssymb,amsmath}
\usepackage[colorlinks,citecolor=red]{hyperref}
\emergencystretch=\maxdimen
\include{MyCommand}

\begin{document}

\title{Principal Component Analysis for Fermionic Critical Points}

\author{Natanael C. Costa}
\email{natanael@if.ufrj.br}
\affiliation{Instituto de Fisica, Universidade Federal do Rio de
Janeiro Cx.P. 68.528, 21941-972 Rio de Janeiro RJ, Brazil}
\affiliation{Department of Physics, University of California Davis, CA 95616, USA}
\author{Wenjian Hu}
\affiliation{Department of Physics, University of California Davis, CA 95616, USA}
\affiliation{Department of Computer Science, University of California Davis, CA 95616, USA}
\author{Z.J. Bai}
\affiliation{Department of Computer Science, University of California Davis, CA 95616, USA}
\author{Richard T. Scalettar}
\affiliation{Department of Physics, University of California Davis, CA 95616, USA}
\author{Rajiv R.P. Singh}
\affiliation{Department of Physics, University of California Davis, CA 95616, USA}


\begin{abstract}
We use determinant Quantum Monte Carlo (DQMC), in combination with the
principal component analysis (PCA) approach to unsupervised learning, to
extract information about phase transitions in several of the most fundamental Hamiltonians describing
strongly correlated materials.
We first explore the zero
temperature antiferromagnet to singlet transition in the Periodic
Anderson Model, the Mott insulating transition in the Hubbard model on a
honeycomb lattice, and the magnetic transition in the 1/6-filled Lieb
lattice.  We then discuss the prospects for learning finite temperature
superconducting transitions in the attractive Hubbard model, for which
there is no sign problem.
Finally, we investigate finite temperature charge density wave (CDW) transitions in the
Holstein model, where the electrons are coupled to phonon degrees of
freedom,
and carry out a finite size scaling analysis to determine $T_c$.
We examine the different behaviors associated with
Hubbard-Stratonovich auxiliary field configurations on both the entire
space-time lattice and on a single imaginary time slice, or other
quantities, such as equal-time Green's and pair-pair correlation
functions.  
\end{abstract}

\maketitle

\section{Introduction}

Deep connections between neural networks, statistical physics, and
biological modeling were established beginning more than three decades
ago.  Hopfield, for example, proposed \cite{hopfield82} a description of
``neurons" whose stable limit points could be used to store specific
``memories", and whose structure was basically that of an Ising model
with symmetric, long range interactions.  A central result was the
emergence of collective properties such as the ability to ``generalize"
to related memories.  Concepts from spin glass theory and frustrated
order were shown to have close analogues with neural networks, including
limitations on the ability to store competing memories; the existence of
critical temperatures for the stability of specific spin patterns
(memories), with additional mixed patterns becoming stable at yet lower
temperature; the role of asymmetric exchange constants (synaptic
strength); and so forth
\cite{amit85a,amit85b,toulouse86,parga86,parisi86,mezard86}.

Over the past several years, the use of neural networks and learning
algorithms has been revisited with fresh perspectives and, in particular
with a focus on the possibility that appropriately defined networks
might be useful in locating phase transitions.
For instance, self-learning algorithms can strongly reduce the autocorrelation from the critical slowing down in second order phase transitions\cite{Liu17,Huang17a,Wang17}.
Furthermore, it was shown \cite{wang16} that PCA could provide a useful diagnostic of the phase
transition in an Ising model in the zero magnetization sector, where two
ferromagnetic domains dominate the partition function at low
temperature.
An extention of this PCA analysis had been done to several other classical
models, e.g.~the Blume-Capel model, which have first order transitions
and tricritical points \cite{hu17}.
Recently, PCA also has been shown relevant for the investigation of the nature
of frustrated classical spin models\cite{WangCe17}.

The well-known exact mapping between the 2D classical Ising model and
the 1D quantum Ising model in a transverse magnetic field immediately
implies that this ability to distinguish a finite temperature transition
in the former case implies that the zero temperature quantum critical
point (QCP) can be determined in the latter.
However, while it is true that all quantum models can be mapped to classical models in one higher
dimension, in most cases the equivalent classical is very complex, for example, for fermions typically depending on a determinant whose entries include the degrees of freedom in the simulation.
It thus remains an open question whether a QCP can be located by learning
methods, although certainly the allowed presence of long range
connections in networks suggests they might be promising.



There have been several recent attempts to combine machine learning
techniques and DQMC.  Ch'ng {\it etal} \cite{ching16} have shown that
convolutional neural networks (CNN) can successfully
{\it generalize} the Ne\'el temperature
$T_N$ of the three dimensional Hubbard model at half-filling.  That is,
a network trained at weak ($U/t=5$) and strong ($U/t=16$) coupling can
predict $T_N$ at intermediate $U/t$, and make inferences concerning the AF
transition when the system is doped (a parameter regime for which the
sign problem prevents direct DQMC simulations at low temperature).
Meanwhile, Broecker {\it etal} \cite{broecker16} have also used DQMC for
the Hubbard model together with CNN with a focus on understanding if the
sign problem can be circumvented.
Learning about the sign problem is
also implicit in machine learning studies of the nodal surfaces of
many-electron wave functions \cite{ledell12}.
A particularly intriguing proposal uses a machine-learned effective bosonic
action to guide proposed moves at a much lower cost than
the usual cube of the system size \cite{xu16,liu16,Huang17b}.

Neural networks have also been used in combination with QMC to provide information of quantum phase transitions\cite{Carrasquilla17,Carleo17,Nieuwenburg17}, topological states\cite{Zhang17a,Torlai17,Zhang17b,Ohtsuki17}, many-body localization \cite{Schindler17,Nieuwenburg17} and entanglement properties\cite{Deng17}.
Furthermore, neural networks are useful to fit functional forms for potential energy surfaces which are then used in subsequent simulations\cite{raff05}.
In contrast to the situations described above, where DQMC is
applied to tight-binding Hamiltonians and the energy scales $U,t,T,\mu$ 
are unambiguously known, 
in these studies the network is used to avoid
complicated and somewhat arbitrary fits to the functional form of the
potential energy, allowing for more robust molecular dynamics
simulations.

Much of the work described above has utilized `supervised' learning
approaches in which the nature of the phases is provided in certain
parameter regimes with the goal of extrapolating the properties of
the patterns to other regimes.
Unsupervised methods have also been used to investigate properties of quantum systems\cite{Ching17,Broecker17}.
Here we study the ability of unsupervised learning methods to determine
the location of the QCP in several models of interacting, itinerant
electrons, when provided with data from DQMC simulations.  Using PCA, we
analyse five cases, including spin, charge and pair ordering.
Concerning magnetic transitions, we investigate (i) the
antiferromagnetic(AF)-singlet transition in the Periodic Anderson Model
(PAM), a Hamiltonian which describes the coupling of a non-interacting
(`metallic') fermionic species with a strongly correlated (`localized')
one; (ii) the paramagnetic metal to insulating antiferromagnet QCP which
occurs at a non-zero $U_c$ in the repulsive Hubbard model on a honeycomb
lattice\cite{paiva05,Herbut06,Assaad13,Parisen15,otsuka16}; and (iii)
the AF ground state for the Hubbard model on the ``Lieb lattice'' at
electronic density $\rho=1/3$, which is closely related to the AF phase
in non-doped cuprates.  We likewise study pair ordering in (iv) the
attractive Hubbard model in the square lattice, which exhibits a finite
temperature phase transition to a $s$-wave superconducting state.  
Finally, charge ordering is investigated in (v) the Holstein model, one of
the simplest Hamiltonians to take into account electron-phonon coupling.
Periodic boundary conditions have been used throughout this paper.
Our methodology focusses on determining
whether the signatures of the transitions occur 
through an examination of the principle components of matrices
constructed from snapshots of the degrees of freedom during the course
of a simulation.

This paper is organized as follow.  The DQMC method and the PCA
procedure are introduced in Section\,\ref{Section:DQMC}.  The PCA
analysis of the PAM is presented in Section\,\ref{Section:PAM}, while the
results for Hubbard model on honeycomb and the Lieb lattices are exhibited
in Sections\,\ref{Section:Honeycomb} and \ref{Section:Lieb},
respectively.  The attractive Hubbard and Holstein models are left to
Sections\,\ref{Section:attHub} and \ref{Section:Holstein}.
The Sections are self-contained, with the individual models 
being briefly introduced in each.

\section{Methodology}\label{Section:DQMC}

DQMC \cite{blankenbecler81} is an approach for solving interacting
fermion Hamiltonians exactly (to within statistical error bars) on
lattices of finite size.  The central observation of the method is that
the partition function ${\cal Z}$ for two fermionic species 
$\sigma = \uparrow,\downarrow$ interacting with a space
and imaginary time dependent bosonic field ${\cal S}(i \tau)$, but {\it
not} with each other, can be written as
\begin{align}\label{eq:partition_func}
{\cal Z} = \sum_{\{S(i \tau)\}}
\, \prod_\sigma {\rm det} 
\big( \,I + B_\sigma(1) B_\sigma(2) B_\sigma(3) \cdots B_\sigma(L) \, \big)
\end{align}
Here the identity matrix $I$ and the matrices $B_\sigma(\tau)$ have dimension
the spatial lattice size $N$, and $L$ is the number of imaginary time
slices into which the inverse temperature $\beta$ is divided.  The sum
is over configurations of the bosonic field.  Each $B_\sigma(\tau)$ is the
product of the exponential of the kinetic energy matrix $K$, which is
usually independent of $\sigma$, and a
diagonal matrix $V_\sigma(\tau)$ whose entries are 
$V_{ii\,\sigma}(\tau) = g_\sigma \lambda {\cal S}(i\tau)$
\footnote{This description of the matrix structure in DQMC is
appropriate for the models studied in this paper.  However, DQMC can
also be used in situations (such as the Su-Schrieffer-Heeger model) in
which the bosonic field modulated the fermion hopping.  In such
situations $K$ becomes $\tau$ dependent and the bosonic field enters its
matrix elements.}.
Here $\lambda$ is the coupling constant between the fermionic and
bosonic variables and $g_\sigma=\pm 1$ depends on the model.  For
Hamiltonians with repulsive interactions, $g_\sigma$ most commonly
has opposite
sign for the two spin species, while for Hamiltonians with attractive 
interactions, $g_\sigma$ is the same for both $\sigma$.
(It is possible to choose $g_\sigma$ to have the same sign for
repulsive interactions, at the expense of introducing an
imaginary coupling constant $\lambda$.)
The separation into exponentials of $K$ and $V$
necessitates an inverse temperature discretization $\Delta \tau=\beta/L$.
This is taken small enough so that systematic `Trotter errors' are
smaller than statistical error bars.  

The Holstein Hamiltonian (see Sec.~VIB) immediately satisfies the
description above.  The field ${\cal S}(i\tau)$ is comprised of the
space-imaginary time values arising from a path integral expression for
the quantum phonon variables.  For the Hubbard model, ${\cal S}(i\tau)$
are the space-time components of a Hubbard-Stratonovich (HS) field
introduced to decouple the fermion-fermion interaction.  In this paper
we employ the discrete version of the HS transformation introduced by
Hirsch \cite{hirsch83}.  Hirsch has shown \cite{hirsch83,hirsch86}
that the correlation functions of the HS variables are directly related
to spin-spin correlations of the fermionic degrees of freedom,
suggesting that success with using magnetic configurations in PCA 
studies of classical transitions\cite{wang16,hu17} might
be replicated with HS configurations.

Principal Component Analysis \cite{Pearson,Wikipedia,Jolliffe} is an
unsupervised learning technique in which, for the implementation here,
configurations of the HS field configurations generated in the course of
a set of DQMC simulations are assembled in the rows of a matrix ${\cal
\bf F}$.  The number of columns of ${\cal \bf F}$ is the dimension of
the HS field (either the spatial lattice size $N$ or the full space-time
lattice size $NL$.  See below.) The number of rows of ${\cal \bf F}$ is
the number of configurations.  Typically we will input $l$
configurations for each of $t$ different simulations which might
correspond to different values of an energy scale in the Hamiltonian,
the temperature, or the density.  Thus the number of rows  of ${\cal \bf
F}$ is $M=lt$.  The mean values of each column of ${\cal \bf F}$ are
subtracted to produce a `data centered' matrix ${\cal \bf X}$.

The most straightforward description of the PCA procedure is that the
eigenproblem of the real symmetric matrix ${\cal \bf X}^T {\cal \bf X}$
is solved, yielding eigenvalues $\lambda_n$ and eigenvectors $w_n$.  The
distribution of the
`relative variances' $\tilde \lambda_n = \lambda_n/ \sum_{i=1}^N
\lambda_i$, and in particular the existence of a gap separating a few
`dominant $\tilde \lambda_n$ from the others, 
provides information about possible phase transitions.  Following the work of 
Wang\cite{wang16}, we will plot the ordered pairs of the first two
`principle components', the inner products of the eigenvectors of 
${\bf X}$ with the two largest eigenvalues, and the {\it individual} HS field
configurations, and also study the `quantified principle
components' which are the {\it averages} of the principle components
over the $l$ configurations of a particular simulation.

The above presentation of PCA has
the virtue of being brief, but does not provide a detailed look at what
the PCA is actually extracting from the data.  For a more complete
exposition see \onlinecite{Pearson,Wikipedia,Jolliffe}.

There are several possible implementations of PCA within the context of
DQMC.  Here we will, for example, examine whether any differences arise between
constructing the PCA matrix ${\bf X}$ from the bosonic field
configuration ${\cal S}(i\tau)$ allowing $i$ to vary over all $N$
spatial sites at a single {\it fixed} $\tau$, as opposed to using ${\cal
S}(i\tau)$ for all $i$ and also all $\tau=1,2,\cdots L$.  
We will also, as previously explored by Broecker {\it etal}
\cite{broecker16} for the repulsive Hubbard model, compare training of a
network with the fermionic Greens function, $G = \big(\, I + B(1) B(2)
\cdots B(L) \, \big)^{-1}$ rather than the HS field configuration.

\begin{figure}[t]
\includegraphics[width=0.98\columnwidth]{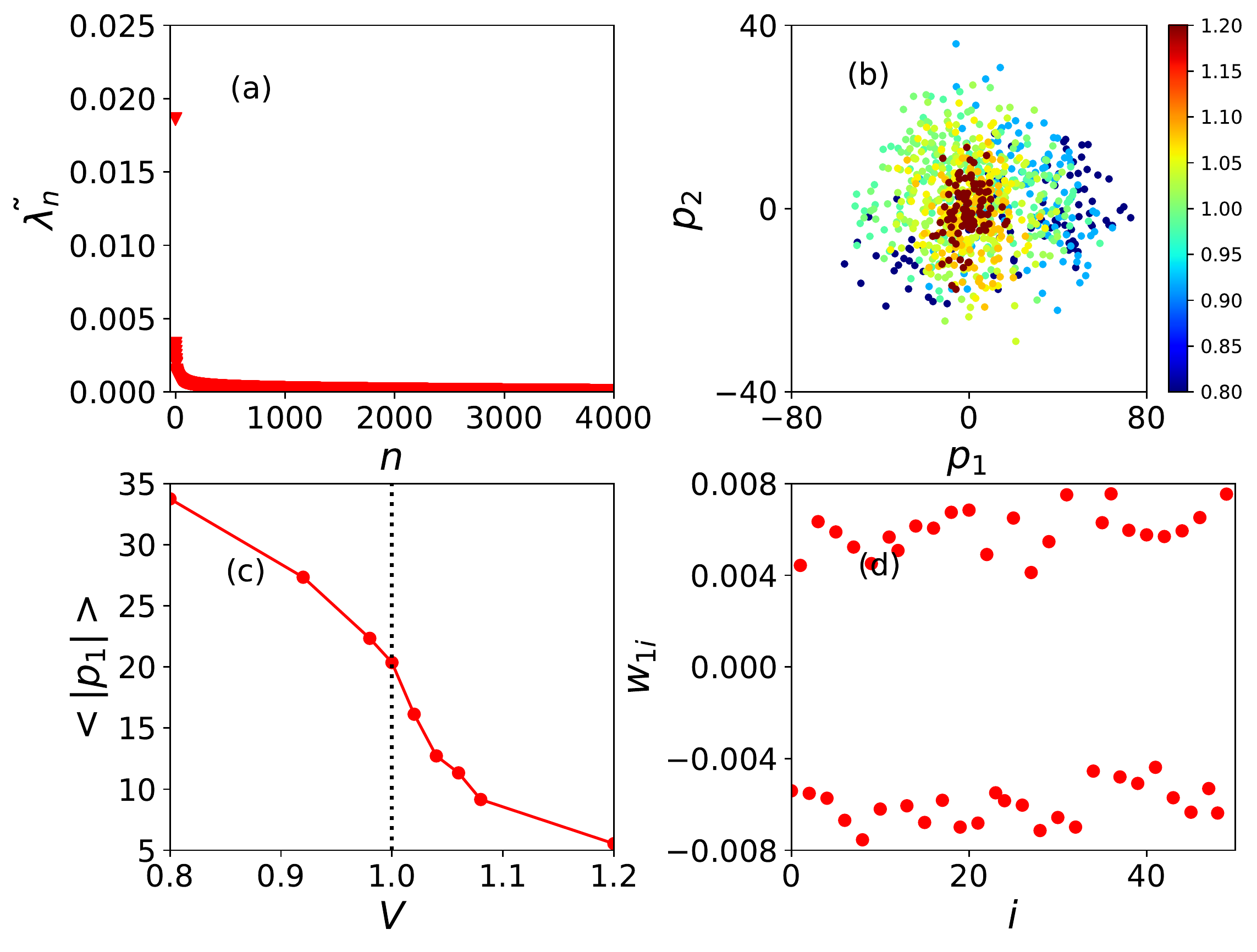} 
\caption{
PCA results for the PAM, with lattice size $N=12 \times 12$, inverse
temperature $\beta/t=24$ and on-site repulsion $U=4$.  (a) Relative
variances $\tilde{\lambda}_n$ obtained from the raw HS field
configurations, with the horizontal axis indicating corresponding
component labels. (b) Projection of the raw HS field configurations onto
the plane of the two leading principal components. Data points are
color-coded by the value of the hybridization $V$ (bar at far right).
For large $V$ (i.e.~in the singlet phase), the pairs form a single small
structure centered at $(0,0)$.  For small $V$ (i.e.~in the AF phase),
the pairs spread out around the origin.  (c) The quantified first
leading component as a function of $V$. The dashed line, corresponding
to the steepest transition, indicates the QCP at $V_c\approx 1.0$.  (d)
The weight vector corresponding to the first leading component, which
shows a clear AF pattern.
\label{fig:PAMeigenvalues}.  
} 
\end{figure}

\section{Results:  AF-Singlet Transition in the PAM}\label{Section:PAM}

We first consider the AF-singlet transition, commonly observed in
heavy-fermion materials~\cite{coleman07},  in which
two species of electrons, conduction ($d$) and localized
($f$) electrons, experience a hybridization $V$ which can be tuned by
adjusting external parameters, such as pressure.  At low $V$,
the Ruderman-Kittel-Kasuya-Yosida (RKKY) interaction leads to long range
magnetic order (LRO), while for large $V$ local singlets form, usurping
LRO.  A quantum phase transition (QPT)
occurs for a
critical hybridization $V_c$ separating these
distinct ground states.  The signature of this QCP is observed even at
finite temperature, with the appearence of non-Fermi liquid behavior.

The standard models for heavy-fermion materials are the Kondo lattice
model (KLM)\cite{doniach77,lacroix79,fazekas91,assaad99,costa17} and the
Periodic Anderson model (PAM)\cite{vekic95,huscroft99,paiva03,hu17b}.
We consider the latter,
which Hamiltonian is
\begin{align}
H = 
-&t \sum_{\langle \mathbf{i},\textbf{j} \rangle \sigma}
\big( d^{\dagger}_{\textbf{i}\sigma}d^{\phantom{\dagger}}_{\textbf{j}\sigma} +
 \mathrm{H.c.} \big)
- V \sum_{\mathbf{i}\sigma} \big( d^{\dagger}_{\textbf{i}\sigma}
f^{\phantom{\dagger}}_{\textbf{i}\sigma}
+ \mathrm{H.c.} \big)
\nonumber
\\
+&U \, \sum_{\mathbf{i}}\big( \, n^f_{\textbf{i}\uparrow}-\frac{1}{2}\,
\big)
\big( \,n^f_{\mathbf{i}\downarrow}-\frac{1}{2}\, \big) \,\,.
\label{eq:hampam}
\end{align}
Here $\mathbf{i}$ runs over sites in a two-dimensional square
lattice, with $\langle \mathbf{i}, \mathbf{j} \rangle$ denoting nearest
neighbors.  $t$ is the hopping integral of conduction $d$-electrons,
and $U$ the on-site Coulomb repulsion in the $f$-band, while their
hybridization is $V$.  The fermion creation (annihilation)
operators of conduction and localized electrons with spin $\sigma$ on a
given site $\mathbf{i}$ are $d^{\dagger}_{{\bf i}\sigma}
(d^{\phantom{\dagger}}_{{\bf i}\sigma})$ and $f^{\dagger}_{{\bf
i}\sigma} (f^{\phantom{\dagger}}_{{\bf i}\sigma})$, respectively.  As
written, the PAM in Eq.\,\eqref{eq:hampam} has particle-hole symmetry
(PHS) so that the density of each electron species obeys $\langle
\rho^d_{{\bf i}\sigma} \rangle =\langle \rho^f_{{\bf i}\sigma} \rangle =
1/2$ at all $t,U,V$ and temperatures.  At this `half-filling', the
AF-singlet QCP occurs at $V_c \approx 0.99 \,t$ for $U=4\,t$ \cite{hu17b}.
Hereafter we set $t=1$ as the scale of energy.

Providing the PCA procedure with the full
space-time HS fields for a simulation on an $12
\times 12$ lattice at $U=4$ and $\beta=24$ (i.e. $L=192$ and $\Delta
\tau=1/8$) for different values of $V$, we obtain the results exhibited
in Fig.~\ref{fig:PAMeigenvalues}.  For each hybridization we provided
$l=1000$ independent configurations.  The relative variance
$\tilde\lambda_n$ for different components $n$ are displayed in
Fig.~\ref{fig:PAMeigenvalues}\,(a), in which the first component is
dominant.  
It has been suggested\cite{hu17} that the appearance of sharp
falloff from dominant relative variances (e.g. $\tilde \lambda_1$) is indicative of a single
dominant spin pattern, e.g.~AF order.
Figure~\ref{fig:PAMeigenvalues}\,(b) presents the projection of the two
leading principal components, with the data points of large
hybridization localized around the origin. For low
hybridization the collection of points bifurcates,
i.e. it appears
two separate clusters indicating the presence of two separate
broken symmetry ground states.
This change is similar to what is seen in finite temperature
transitions in classical spin models.
Here we interpret the analogous behavior as
signaling a QPT at a critical $V_{c}$.

The position of the QCP can be roughly inferred via the behavior of the
quantified first leading component as a function of $V$, as displayed in
Fig.~\ref{fig:PAMeigenvalues}\,(c).  For low hybridizations $\langle |
p_{1} | \rangle$ is large, while it is suppressed at large
hybridizations, behaving similar to the AF structure factor.  One can
estimate the QCP location at the inflection point, $V_{c} \approx
1$, where the $\langle | p_{1} | \rangle$ is most rapidly changing.
This $V_{c}$ is in agreement with conventional approaches based on
finite size scaling of the AF structure factor\cite{vekic95,hu17b},
$V_c \sim 0.99\,t$,
although at present the PCA analysis is clearly considerably less accurate.
That order at wavevector ${\bf q}=(\pi,\pi)$ is demonstrated by the structure of the first 
eigenvector, Fig.~\ref{fig:PAMeigenvalues}\,(d), which has an obvious alternation in sign for sites on the two sublattices.
In classical transitions\cite{hu17}, the quantified leading
components mimic physical quantities such as magnetization and susceptibility. 
Interestingly, the quantified first leading component of the PAM also resembles the mean 
field magnetization, as shown in Fig.\,2 of the Ref.\,\onlinecite{hu17b}.

Notably, it has been suggested\cite{hu17} for classical
models that the number of distinct groupings
of the $(p_1,p_2)$ distributions reflects the degeneracy of the ground
state: two groups in the Ising case where spins can order either
up or down, a continuous ring around the origin for the XY model, 
and four peaks for a biquadratic Ising model which possesses four 
ordered phases.
The AF order in the PAM, and in the Hubbard and Lieb models
below, has a continuous symmetry similar to the XY case.
However, the HS transformation used here, and
in many DQMC simulations, couples preferentially to the $z$ 
component of the fermion spin.  This choice does {\it not}
represent any approximation in the results for physical
observables obtained
by DQMC, but is known to break symmetries in quantities
like {\it error bars} which are algorithm-dependent.
The $(p_1,p_2)$ distributions, shown in Fig.~\ref{fig:PAMeigenvalues}\,(b), 
are more symmetrically distributed about $(0,0)$ than for
the Ising model\cite{wang16}, 
because the system size is still relative small and the temperature is low but not zero.

If a rotationally invariant HS decoupling is performed,
and the resulting configurations are fed into the PCA, 
symmetry is preserved and similar results to the XY model are expected. 
There are very significant drawbacks to this approach.
The two spin                                                                   
species are mixed by coupling of a HS field $S_x$ to the $x$
component of the fermion spin, 
i.e.~$S_x \, ( \, c^\dagger_{i\uparrow} c^{\phantom{\dagger}}_{i\downarrow}
+ c^\dagger_{i\downarrow} c^{\phantom{\dagger}}_{i\uparrow}) $.
As a consequence, the two independent
spin up and spin down
Greens functions (and fermion determinants) become a single matrix with
double
the dimension.  This fundamental change in the algorithm,
which several groups have attempted, 
including ourselves\cite{batrouni90,chen92},
is known (in the case of the Hubbard model) to worsen the sign problem.
In the present case of the Holstein model, there is no sign problem
precisely because there are two separate determinants whose
square is always positive. In addition, execution time is significantly
increased by the necessity of inversion of a larger matrix.

\section{Results: AF-PM Transition on a Honeycomb Lattice}\label{Section:Honeycomb}

We next investigate the Hubbard Hamiltonian, 
\begin{align}
H = -&t \sum_{<\textbf{i},\textbf{j}>\sigma}
\big( d^{\dagger}_{\textbf{i}\sigma}d^{\phantom{\dagger}}_{\textbf{j}\sigma}+
d^{\dagger}_{\textbf{j}\sigma}d^{\phantom{\dagger}}_{\textbf{i}\sigma}\big)
\nonumber
\\
+&U\sum_{\textbf{i}}\big( \, n^d_{\textbf{i}\uparrow}-\frac{1}{2}\,
\big)
\big( \,n^d_{\textbf{i}\downarrow}-\frac{1}{2}\, \big) ,
\label{eq:hamhub}
\end{align}
on a honeycomb lattice.  Unlike the square lattice, in this geometry
a critical value $U_{c}$ is required in order to obtain an AF ground
state at half filling\cite{paiva05,Herbut06,Assaad13,Parisen15,otsuka16}.  A
metal-insulator transition also occurs at $U_{c}$, in
contrast with the PAM, where both AF and singlet phases are insulators.

\begin{figure}[t]
\includegraphics[width=0.98\columnwidth]{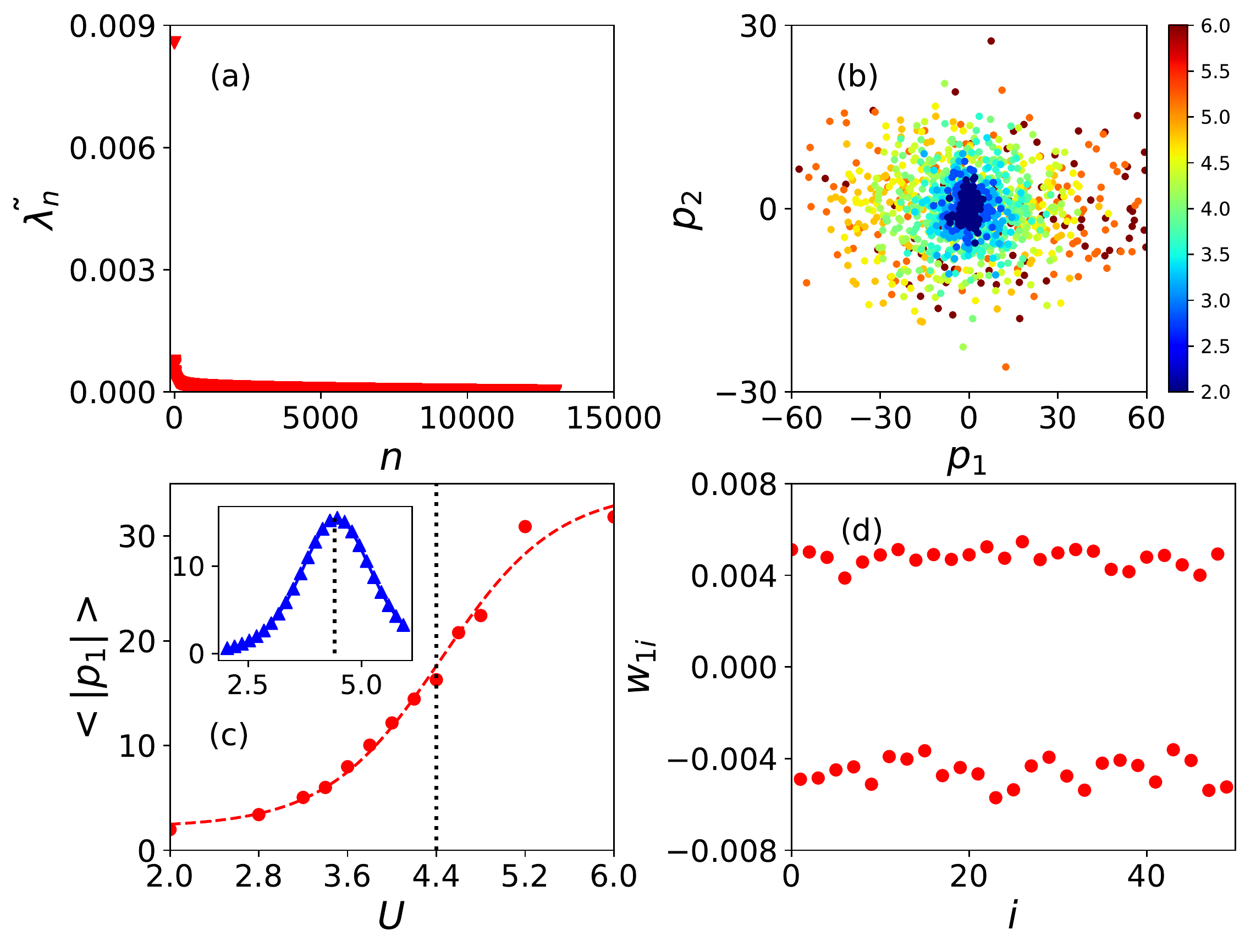} 
\caption{
PCA results for the honeycomb lattice (single band) Hubbard model, with
the lattice size $N=12 \times 12$, the inverse temperature $\beta/t=20$.
(a) Relative variances $\tilde{\lambda}_n$ obtained from the raw HS
field configurations. (b) Projection of the raw HS field configurations
onto the plane of the two leading principal components. Data points are
color-coded by the value of the onsite repulsion $U$ (bar at far right).
For small $U$ (i.e.~in the paramagnetic phase), the pairs are localized
near $(0,0)$.  For larger $U$ (i.e.~in the AF phase), the pairs spread
out around the origin. (c)
The symbols corresponds to the quantified first leading component as a
function of $U$, while the dash (red) curve is just a guide to the eyes.
The vertical (black) dotted line, corresponding to the steepest
transition, indicates the QCP at $U_c\approx 4.4$.  
The inset displays the numerical derivative.
(d) The weight
vector corresponding to the first leading component, which shows a clear
AF pattern.
\label{fig:honeyeigenvalues} } 
\end{figure}
Here we examine the use of the PCA to discern the finite $U_c$ required to
induce AF order.  We performed simulations on a $12 \times 12$ lattice
at $\beta=20$ (i.e. $L=160$ and $\Delta
\tau=1/8$), and providing $l=1000$ independent configurations for each
value of $U$.  In Fig.~\ref{fig:honeyeigenvalues}\,(a) we observe, as
for the PAM, a single dominant relative variance.  The pairs of the two
largest components are shown in
Fig.~\ref{fig:honeyeigenvalues}\,(b), and provide information about
the QCP: A group of points centered around the origin for small $U$
spreads out rapidly for large $U$.  An estimation of the QCP is
obtained from the quantified first leading component as a function of $U$,
exhibited in Fig.~\ref{fig:honeyeigenvalues}\,(c).
For a better determination of the inflection point one can perform a
numerical fitting of the DQMC data, with
its differentiation providing a maximal value for
$d\langle |\,p_1\,|\rangle/dU$
at $U \approx 4.4$,
as displayed in the inset of Fig.~\ref{fig:honeyeigenvalues}\,(c).
This result is reasonably close to the critical value
obtained by conventional methods, where $U_c \approx 3.85$\cite{otsuka16}.
However, the evolution of $\langle |\,p_1\,|\rangle$ is quite
gradual. 
The estimation of $U_{c}$ will likely be improved
with the analysis of different lattice sizes\citep{hu17};
e.g. see Section\,\ref{Section:Holstein}.
It appears the determination of the QCP is less accurate
using the PCA than from conventional scaling methods on lattices
of the same size, as was also seen in the preceding Section
for the PAM. 
Examination of the eigenvector
[Fig.~\ref{fig:honeyeigenvalues}\,(d)] reveals a staggered pattern
which indicates the ordering is AF.

\begin{figure*}[t]
\includegraphics[width=0.8\textwidth]{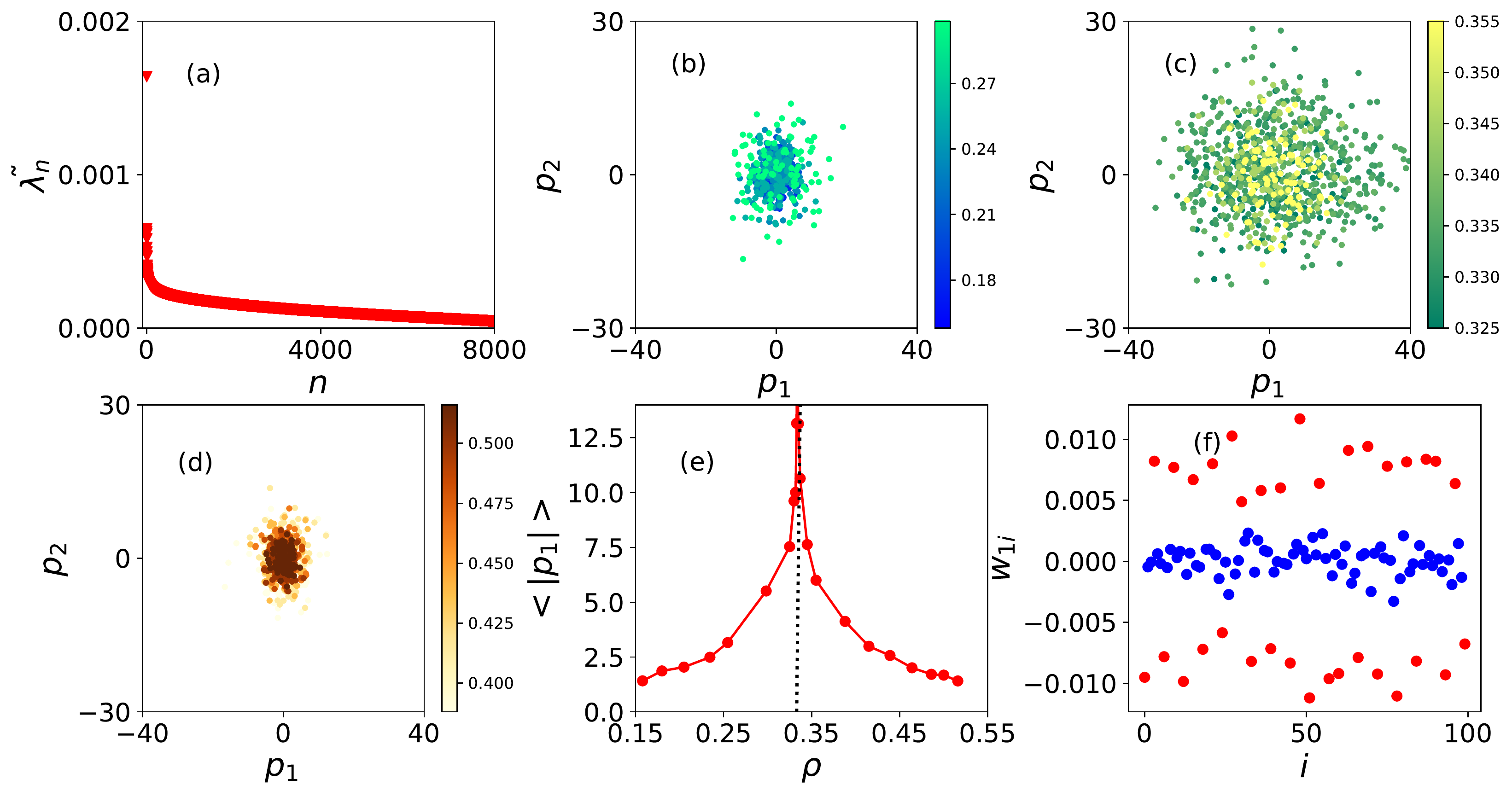} 
\caption{
PCA results for the Lieb lattice Hubbard model, with lattice size
$N=10 \times 10$, and inverse temperature $\beta/t=20$.  (a) Relative
variances $\tilde{\lambda}_n$ obtained from the raw HS field
configurations. (b)-(d) Projection of the raw HS field configurations
onto the plane of the two leading principal components. Data points are
color-coded by the value of the onsite density $\rho$ (bar at far
right).  (e) The quantified first leading component as a function of
$\rho$. The dashed line, showing the abrupt peak, indicates the QCP at
$\rho_c\approx 1/3$, (f) The weight vector corresponding to the first
leading component. Red dots represent (copper) $d$-sites, while blue
dots correspond to (oxygen) $p$-sites. 
\label{fig:liebeigenvalues} } 
\end{figure*}

\section{Results: Lieb Lattice at 1/6 Filling}\label{Section:Lieb}

As a final illustration of magnetic transitions in the ground state,
we study the
repulsive Hubbard Hamiltonian on the ``Lieb lattice".  In contrast to
the two previous cases, where energy scales $V$ and $U$ are used to tune 
through
the QCP, the transition is explored as a function of filling.
The Lieb lattice describes a three
band model formed by an underlying square lattice which is then
decorated with additional sites on each bond.  The Lieb lattice is
bipartite, but has a different number of sites on the two sublattices.
An interesting feature of this geometry is the presence,
in the noninteracting limit, of a perfectly
flat energy band, sandwiched between two dispersing bands,
which can lead to ferromagnetism when $U$ is turned on
at half filling\cite{vollhardt01}.

In a seminal paper, Lieb showed that the Hubbard model on this lattice
at half filling exhibits long range ferrimagnetic order in
its ground state\cite{lieb89}.  The spin order in similar geometries
has been explored in
Refs.\,\onlinecite{shen94,mielke91,tasaki95,tasaki98,tasaki98b}.  In
addition to these rigorous results, the Lieb lattice is of 
interest as a more faithful representation of the CuO$_2$ sheets of
cuprate superconductors than is provided by the single band Hubbard
model.  In this three band case, the repulsion $U$ is
typically chosen to take different values on the square lattice and
bridging sites.  The implications of inhomogeneous $U$ for
ferrimagnetism were recently explored in Ref.\,\onlinecite{costa16}.

Lieb's theorem is of limited direct interest for cuprate
superconductivity since it describes a filling of the lattice $n=3$
holes per CuO$_2$ unit cell, far from
that actually present in these materials,
$n = 1+\delta$.
Although there are no rigorous theorems available for the nature of the
spin order away from half filling, previous results provide evidence of
antiferromagnetic correlations for one hole per unit cell, which
are strongly suppressed for small doping $\delta$
\cite{scalettar91,huang01,kung16}.  The investigation of
other phases, such as superconductivity for small doping, remains a
challenge.

Here we apply PCA to determine the enhancement of
antiferromagnetic correlations for Hubbard model in the Lieb lattice at
fillings around one hole per unit cell (or $\rho=1/3$).  
Away from half filling, the measurements of physical
quantities by DQMC are strongly impeded by the sign problem, see
e.g.~Ref.\,\onlinecite{santos03}.  Since the PCA
procedure can be provided with HS fields without the necessity of
dividing by the average sign, it can be undertaken even when the
average sign is small.  It should be noted, however, that even in this
case the HS fields are generated with the absolute value of the
determinants, and it remains an open question how much this 
will bias the physics\cite{white89}.  
We return to this point in the conclusions.

The Hubbard Hamiltonian on the Lieb lattice is
\begin{align}
H = &-t_{pd} \sum_{{\bf r} \sigma}
\big( \,
d_{{\bf r} \sigma}^{\dagger}   \, p_{{\bf r} \sigma}^{x\phantom{\dagger}}
+ d_{{\bf r} \sigma}^{\dagger}   \, p_{{\bf r} \sigma}^{y\phantom{\dagger}}
+ {\rm h.c.} \big)
\nonumber \\
&-t_{pd} \sum_{{\bf r} \sigma}
\big( \,
d_{{\bf r}  \sigma}^{\dagger}   
\, p_{{\bf r} -\hat x \, \sigma}^{x\phantom{\dagger}}
+ d_{{\bf r}  \sigma}^{\dagger}   
\, p_{{\bf r} -\hat y \, \sigma}^{y\phantom{\dagger}}
+ {\rm h.c.} \big)
\nonumber \\
 &+  \sum_{{\bf r} \alpha} U_\alpha
 \left(\, n^{\alpha}_{{\bf r} \uparrow} -\frac12 \, \right)
 \left(\, n^{\alpha}_{{\bf r} \downarrow } -\frac12 \, \right)
\nonumber \\
&
- \mu  \sum_{{\bf r} \alpha \sigma}  n^{\alpha}_{{\bf r} \sigma} 
\label{eq:Lieb_hamil}
\end{align}
with $t_{pd}$ being 
hopping between $d$ and
$p^\alpha$ ($\alpha=x$ or $y$) orbitals.
As before
we define $t_{pd}$ as unity.
We investigate the inhomogeneous on-site repulsion case, with $U_p=0$
and $ U_d = 4$.  For this choice of $U_\alpha$, the particle-hole
symmetric form of the interaction energy in 
Eq.\,\eqref{eq:Lieb_hamil} leads to a difference of on-site energies
$\Delta \varepsilon_{dp} = 2$, which is close to the difference of
on-site energies of oxygen and copper orbitals in the cuprates.

\begin{figure*}[t]
\includegraphics[width=0.8\textwidth]{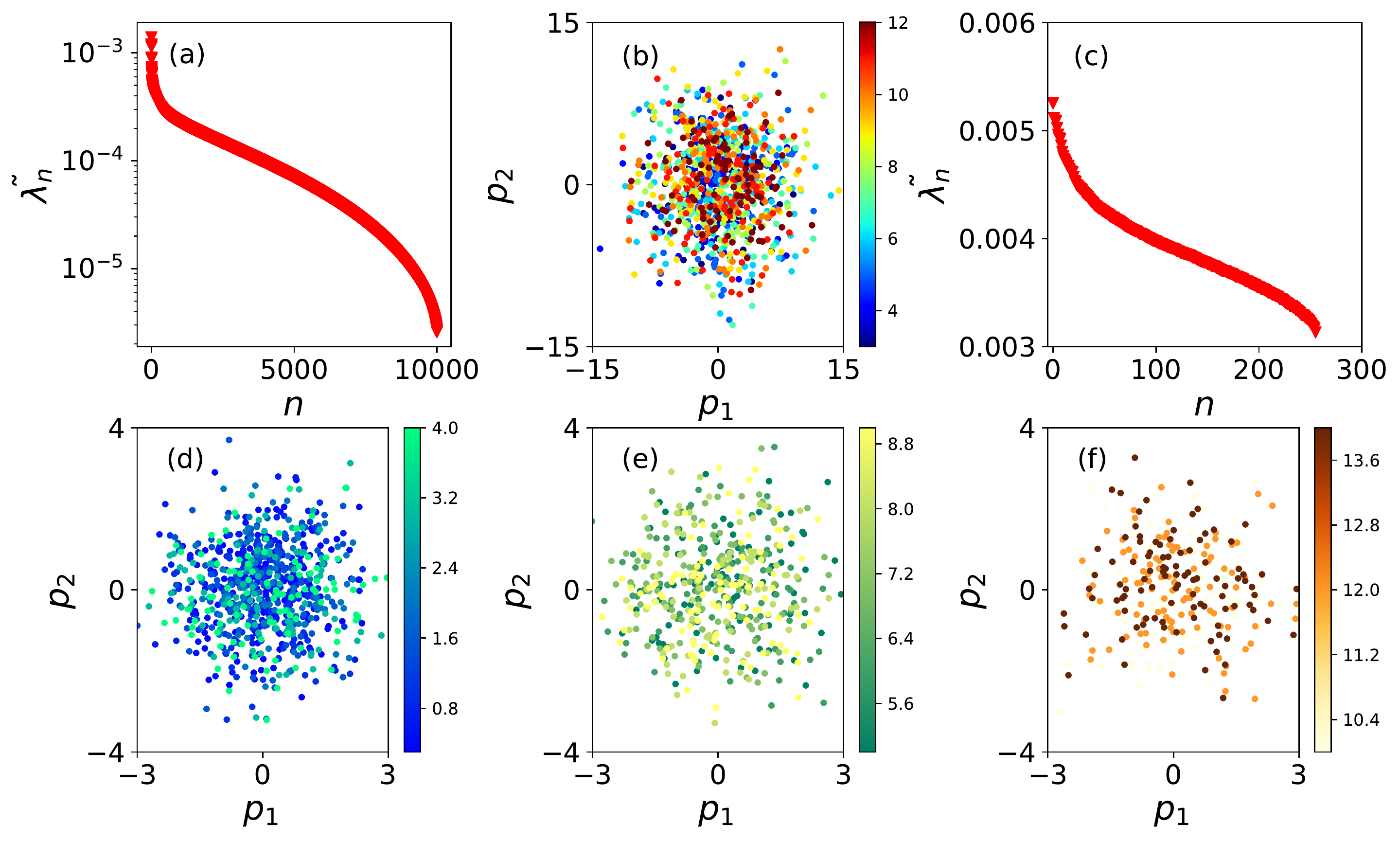} 
\caption{Results of PCA analysis of the attractive 
Hubbard model.  In panels (a,b) the full space-time HS field
variables were used and $\beta$ was altered over the
range given by the color bar to the right of (b) at fixed $L$ by
varying $\Delta \tau$.  The spatial lattice size 
was $12 \times 12$.
In panels (c,d,e,f) the spatial lattice size was $16 \times 16$ and only
the HS variables on a single slice were given to the PCA. 
In the latter case, for clarity, the $(p_1,p_2)$ distribution is
further separated into three panels.
}
\label{fig:aHubbardHSF}
\end{figure*}

We performed simulations on a $10 \times 10$ lattice (i.e. 300 sites) at
$\beta=20$ (fixing $L=160$ and $\Delta \tau = 1/8$), and providing $l=1000$ independent configurations for each value
of $\rho$.  The relative variances $\tilde\lambda_n$ for different
components $n$ are displayed in Fig.~\ref{fig:liebeigenvalues}\,(a).
Although the sub-dominant values are more prominent than in the PAM and
Hubbard model cases, $\lambda_1$ is still more than twice $\lambda_2$.
The projection of the two largest components, is presented in
Figs.~\ref{fig:liebeigenvalues}\,(b)-(d), for (b) $\rho < 1/3$, (c)
$\rho \sim 1/3$ and (d) $\rho > 1/3$.  Notice that for $\rho < 1/3$ and
$\rho > 1/3$ the data points form a small cluster around the origin,
whereas for $\rho \sim 1/3$ they are spread out, suggesting a
disorder-order-disorder transition when $\rho$ varies.
Figure \ref{fig:liebeigenvalues}\,(e) exhibits the quantified first
leading component as a function of $\rho$, with a sharp increase in
$\langle | p_{1} | \rangle$ occuring at $\rho =1/3$.  A direct
comparison can be made between the $\langle | p_{1} | \rangle$ and the
AF structure factor of $d$-sites, which provides evidence of an AF
ground state at $\rho =1/3$, in line with the conventional
analysis\cite{scalettar91,huang01,kung16}.

Figure \ref{fig:liebeigenvalues}\,(f), the leading eigenvector, 
emphasizes that the magnetic order is on the `copper' sites
of the square sublattice, while the `oxygen' bridging sites
have nearly zero components.

\color{black}
\section{Results: Superconductivity in Attractive Hubbard Model} \label{Section:attHub}

The previous Sections have described the ability of PCA to learn about
``quantum critical points"
as energy scales in the Hamiltonian or
density are varied.  We now turn our attention to examining the finite
temperature transition in the attractive Hubbard model.

At half-filling, and on a bipartite lattice, a particle-hole
transformation (PHT) on the Hubbard Hamiltonian maps the attractive to
the repulsive cases, so that the existence of AF order at $T=0$ in the
latter implies the presence of simultaneous CDW (the analog of AF in the
$z$ direction) and Superconductivity (SC) order (the analog of AF order in the $xy$ plane)
in the former \cite{scalettar89}.  Thus the results of
Sec.\,\ref{Section:Honeycomb} for the AF transition of the 
repulsive Hubbard model immediately imply 
that PCA can capture the SC-CDW transition
in the ground state of the half-filled attractive Hubbard model on a
honeycomb lattice.  Away from half-filling it is known from the PHT,
which maps the model to the repulsive model in an external Zeeman field,
that there is a finite temperature Kosterlitz-Thouless (KT)
transition\cite{scalettar89} to a purely SC state.
For instance, at $\rho=0.80$ on a
square lattice one obtains $T_{c} \approx 0.13$,
as reported by Ref.\,\onlinecite{paiva04}.

Earlier work  on the classical XY model \cite{hu17} suggests PCA can
capture aspects of KT physics -- the presence of a transition --
but not its physical origin in terms of vortex unbinding.
We now use the method to
investigate KT physics in a quantum Hamiltonian.  
The analysis of finite temperature transitions in the PCA approach is
complicated by the fact that varying $T$ changes the number of imaginary
time slices at fixed $\Delta \tau$, and, therefore, the number of HS
fields.  Thus the number of columns of ${\bf X}$ varies over the
$t$ simulations.  There are two possible solutions.  One is to sweep the
temperature by varying $\Delta \tau$ at fixed $L$.  (If this is done, one
must ensure $\Delta \tau$ remains small enough throughout is variation
so that Trotter errors are minor.)  The other approach is to give PCA
only the HS variables 
on a single time slice.  One might expect the PCA to be able to
discern transitions even when not given the imaginary time evolution,
since one knows that conventional approaches can analyze magnetic order
both by using only equal-time measurements (the structure factor) or
by analyzing unequal-time measurements (the susceptibility).

Figure \ref{fig:aHubbardHSF} shows the results from both approaches, in which we fixed $U=-4$ in Eq.\,\eqref{eq:hamhub}, and adjusted $\mu$ to have $\rho=0.80$.
In panels (a) and (b) the full space-time HS variables of a $12 \times 12$ square lattice are given to the PCA.
The inverse temperature $\beta$ is changed by altering $\Delta \tau=[0.03, 0.12]$ at fixed $L=100$.
We provided $l=1000$ independent configurations for each temperature to our PCA procedure.
Panels (c)-(f) are concerning to single time slice measurements in a $16 \times 16$ square lattice at fixed $\Delta \tau=0.125$ and varying $L$.
For this latter case we provided $l=4000$ independent measurements to PCA.

The PCA does not appear to capture the finite $T$ SC transition 
in the attractive Hubbard model:
There appears to be very little change in the geometry of the
$(p_1,p_2)$ distribution in going through $T_c$.
This absence of a signal is true both when the full space-time field is provided
[Fig.~\ref{fig:aHubbardHSF}\,(b)] and also when just the spatial components are given
[Fig.~\ref{fig:aHubbardHSF}\,(d)-(f)].
We also do not observe any relevant variation in the quantified first
leading component as a function of $\beta$.  
A possible origin of this failure is that the HS field couples to
the spin order, so that while it can, as seen earlier,
carry information to the PCA about magnetic transitions,
it is not able to do so for pairing transitions.
In principle it is possible to use HS fields which couple
to the pair creation and destruction operators,
but this transformation results in a very bad sign problem,
even for the attractive model\cite{batrouni90}.

\begin{figure}[t]
\includegraphics[width=0.98\columnwidth]{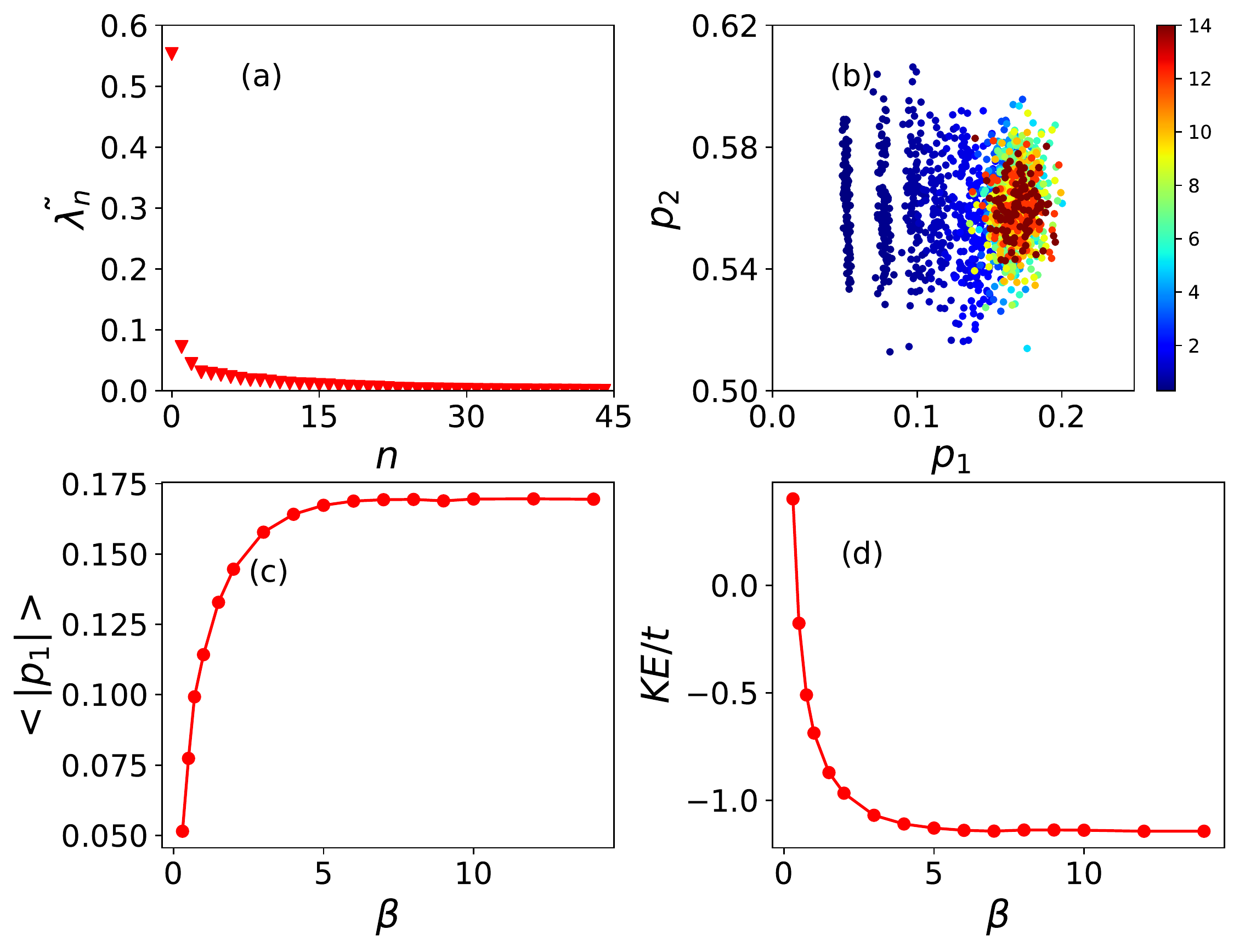} 
\caption{
PCA results for the attractive Hubbard model with lattice size
$N=16 \times 16$, onsite repulsion $U=-4$ and density $\rho=0.8$.
(a) Relative variances $\tilde{\lambda}_n$ obtained from the raw Green
functions.  (b) Projection of the raw Green functions onto the plane of
the two leading principal components.  The color bar indicates the
inverse temperature $\beta$ in units of $t$.  (c) The quantified
first leading component as a function of $\beta$. (d) The kinetic energy
(KE) as a function of $\beta$.
\label{fig:aHubbardeigenvalues_1} } 
\end{figure}

Having seen inconclusive results when providing the PCA with the HS
field configurations, we follow recent work\cite{broecker16} and employ
the equal-time Green's functions, $G_{\mathbf{i}\mathbf{j}}$.  In
contrast to Fig.~\ref{fig:aHubbardHSF}\,(a), the relative
variances now exhibit a single dominant component, as seen in
Fig.\,\ref{fig:aHubbardeigenvalues_1}\,(a).  Furthermore,
Fig.\,\ref{fig:aHubbardeigenvalues_1}\,(b), data for the two largest
components distinguish high and low temperatures.  The quantified first
leading component increases in magnitude as $\beta$ increases
[Fig.\,\ref{fig:aHubbardeigenvalues_1}\,(c)], showing a very similar behavior
to the kinetic energy (KE) of the system, exhibited in
Fig.\,\ref{fig:aHubbardeigenvalues_1}\,(d).  This connection is not
surprising.  $G_{\mathbf{i}\mathbf{j}}$ is closely related to the
mobility of the electrons in the system, enabling the PCA to discern
different regimes.  However, the
method does not capture the SC phase transition in the attractive
Hubbard model, in the sense of showing a definitive signal near $\beta_c
= 1/T_c \sim 7$.  A similar smooth evolution of the kinetic energy
(``effective hopping") is seen in the half-filled repulsive Hubbard
model as $U$ is increased\cite{white89}.

\begin{figure}[t]
\includegraphics[width=0.98\columnwidth]{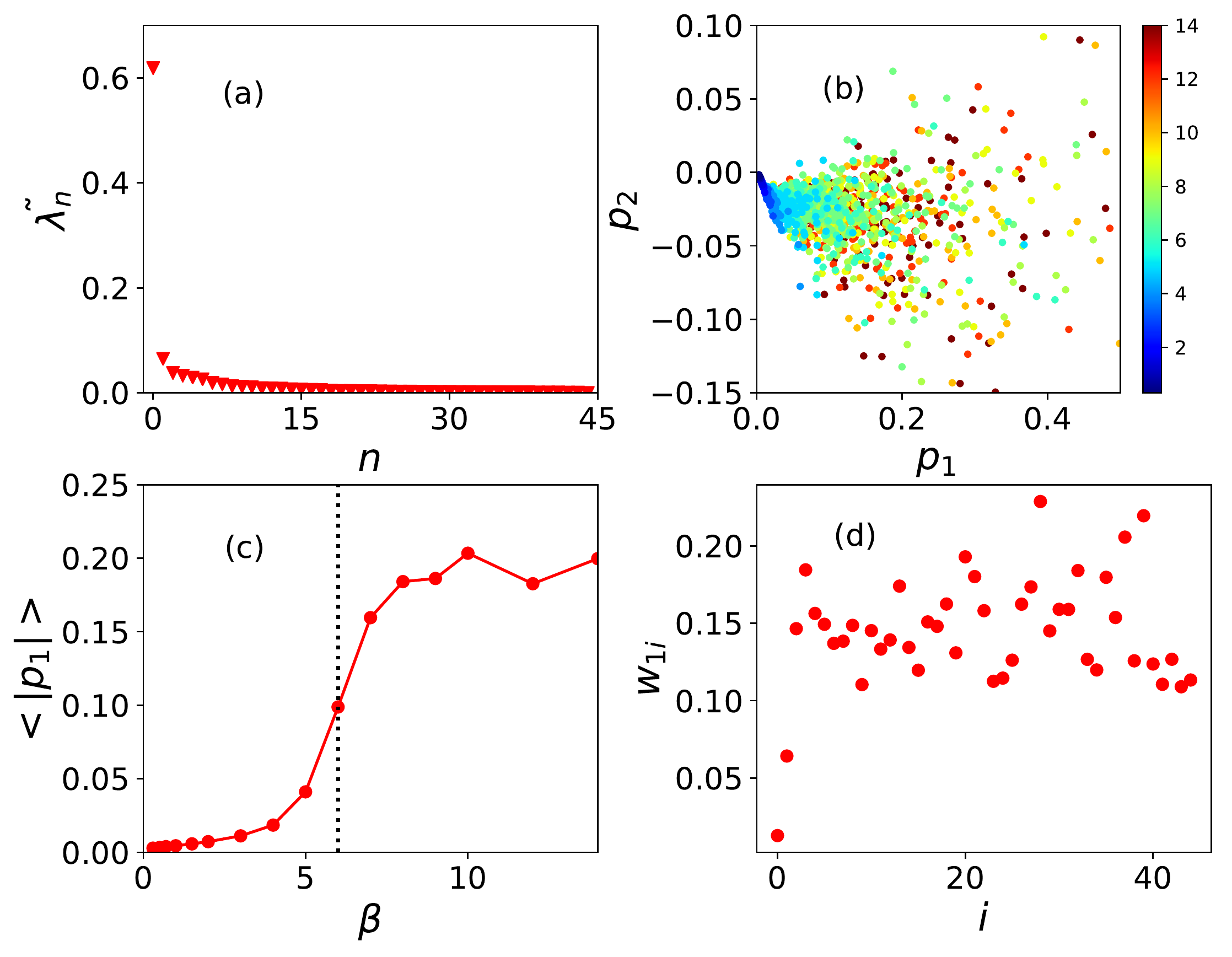} 
\caption{
PCA results for the attractive Hubbard model with the lattice size
$N=16\times 16$, the onsite repulsion $U=4$ and the density $\rho=0.8$.
(a) Relative variances $\tilde{\lambda}_n$ obtained from the raw pair
pair correlation function. (b) Projection onto the plane of the two leading principal
components.  The color bar indicates the inverse temperature $\beta$ in units of $t$.  (c) The quantified first leading component as a
function of $\beta$. The dashed line, corresponding to the steepest
transition, indicates the QCP at $\beta_c\approx 6.0$.  (d) The weight
vector corresponding to the first leading component.
\label{fig:aHubbardeigenvalues_2} } 
\end{figure}

As a final attempt to use the PCA to observe the SC transition
in the attractive Hubbard model, we use the equal-time pair-pair
correlation functions, 
\begin{align}
\Gamma_{\mathbf{i} \mathbf{j}} = \langle d^{\dagger}_{\mathbf{i} \uparrow} d^{\dagger}_{\mathbf{i} \downarrow} d_{\mathbf{j} \downarrow} d_{\mathbf{j} \uparrow} + \mathrm{H.c.} \rangle .
\end{align}
Figure \ref{fig:aHubbardeigenvalues_2}\,(a) displays the relative variances
of the principal components, whereas
Fig.~\ref{fig:aHubbardeigenvalues_2}\,(b) exhibits the projection of the two
largest components.  The former presents a single dominant component,
while the later shows two different behaviors to low and high
temperatures.  As before, we analyse the quantified first leading
component $\langle | p_{1} | \rangle$ as a function of $\beta$.
This is seen to 
behave in a similar way as the uniform Fourier transformation of the
pair-pair correlation functions, $P_{s}$, as displayed in
Fig.~\ref{fig:aHubbardeigenvalues_2}\,(c).  These
allow the PCA to provide the most promising signal of
the SC phase transition around $\beta_{c}=6$, in rough agreement
with the known critical temperature $\beta_{c} \approx 7.5$
\cite{paiva04}.  The conventional approaches which yield this value
involve a demanding process of data collapses of $P_{s}$, a level of
analysis which this initial PCA study here cannot attempt, since data on
only a single lattice size is studied.

Unlike previous models where the ordering vector is $(\pi,\pi)$,
the pairing amplitude is uniform in the attractive Hubbard model.
This is reflected in the lack of oscillations in the principal
eigenvector, Fig.~\ref{fig:aHubbardeigenvalues_2}\,(d).

\section{Results: Charge-density wave in Holstein Model}\label{Section:Holstein}

Finally, we study the finite temperature CDW transition in 
the Holstein model\cite{Holstein59}, one of the simplest tight-binding
models of the electron-phonon interaction (EPI).  
The Holstein model describes
independent (i.e.~dispersionless) quantum harmonic oscillators (HO) 
interacting locally with the electron density,
\begin{align} \label{eq:Holst_hamil}
\nonumber \mathcal{H} = &
-t \sum_{\langle \mathbf{i}, \mathbf{j} \rangle, \sigma} \big(d^{\dagger}_{\mathbf{i} \sigma} d_{\mathbf{j} \sigma} + {\rm h.c.} \big)
- \mu \sum_{\mathbf{i}, \sigma} n_{\mathbf{i}, \sigma}
- \lambda \sum_{\mathbf{i}, \sigma} n_{\mathbf{i}, \sigma} \hat{X}_{\mathbf{i}}
\\
&
+ \frac{1}{2} \sum_{ \mathbf{i} } \hat{P}^{2}_{\mathbf{i}}
+ \frac{\omega_{0}^{2}}{2} \sum_{ \mathbf{i} } \hat{X}^{2}_{\mathbf{i}}
,
\end{align}
As earlier, the sum over $\langle \mathbf{ij} \rangle$ is over nearest
neighbor sites on a two-dimensional square
lattice.  $\hat{P}$ and $\hat{X}$ are respectively the momentum and
displacement operators of HOs with frequency $\omega_{0}$ and mass
$m=1$.  The electron-phonon coupling is $\lambda$ which, when integrated
out, neglecting the $\hat{P}^2$ terms, leads to a 
dimensionless coupling, 
$\lambda_{D}= \lambda^2 \,/\,2 t \omega^2_{0}$. 

The Holstein Hamiltonian is quadratic in the fermion
operators, which can therefore be integrated out without the 
introduction of a HS
field.  The partition function then involves an integration over the
phonon degrees of freedom,
\begin{align}\label{eq:partition_func_Hols}
Z = \int \mathrm{d}\{x_{i,l}\} e^{-\Delta \tau S_{B}} 
\bigg[ \mathrm{det}\big(I + B_{1} B_{2} \cdots B_{L} \big) \bigg]^2,
\end{align}
with $\int \mathrm{d}\{x_{i,l}\}$ being the integral over the set of
continuous variables $x_{i,l}$ and 
\begin{align}
S_{B} = \sum_{i=1}^{N}
\sum_{l=1}^{L} \bigg[ \frac{1}{2m} \bigg( \frac{x_{i,l} -
x_{i,l+1}}{\Delta \tau} \bigg)^2 + \frac{m \omega_{0}^{2}}{2}
x^{2}_{i,l} \bigg]
\end{align}
the phonon action.  
Because the phonons couple to the charge, symmetrically for the spin up
and spin down species, $g_\sigma$ is the same for
$\sigma=\uparrow,\downarrow$ and the two determinants are identical.
Their
product is always positive and there is no sign problem for any filling
(as is also the case for the attractive Hubbard model).  The phonon
fields $\{x_{i,l}\}$ are sampled by standard Monte Carlo.

\begin{figure}[t]
\includegraphics[width=0.98\columnwidth]{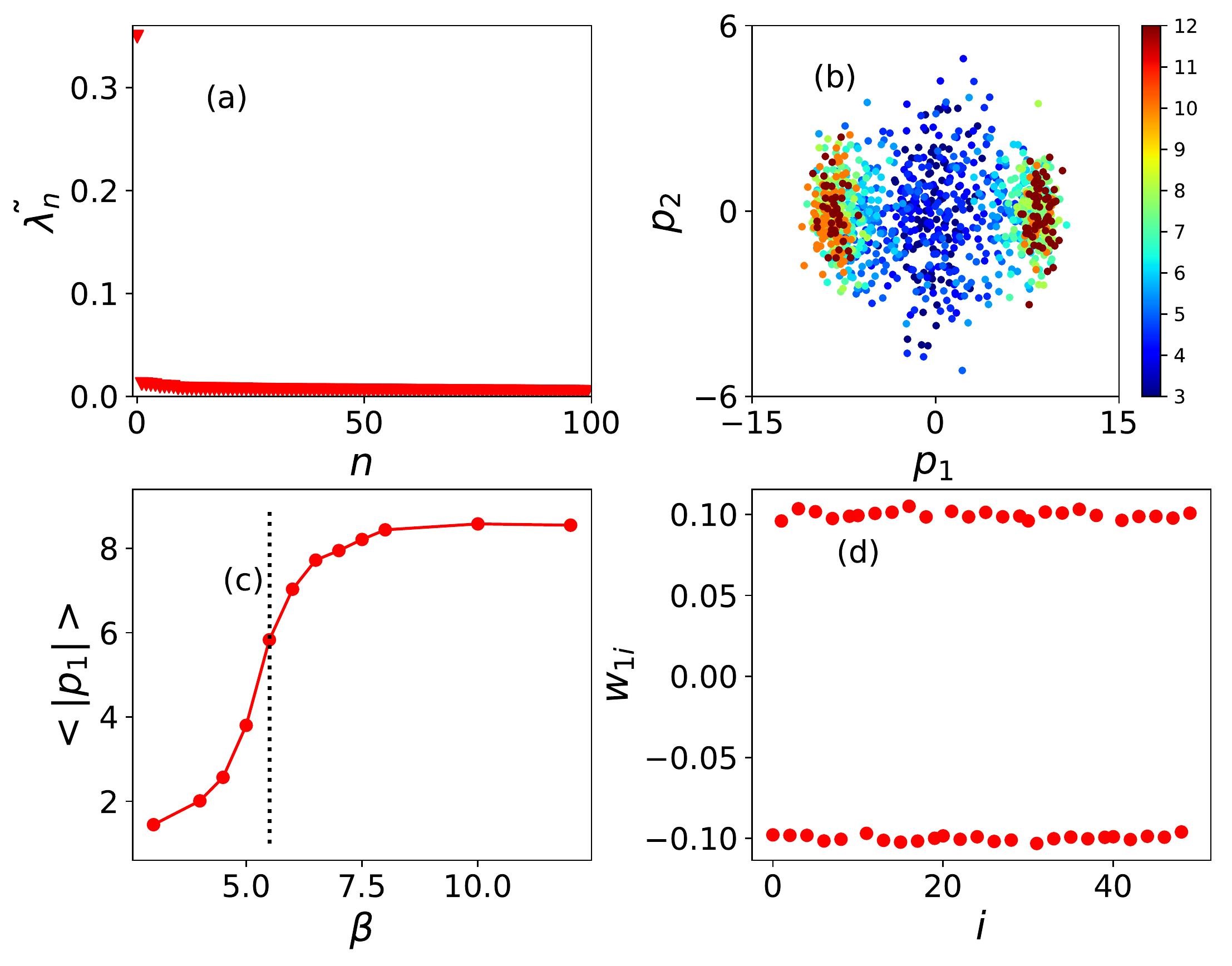} 
\caption{
PCA results for the half-filled $10\times10$ Holstein model for $\lambda_{D}=\omega_{0}=1$. 
(a) Relative variances $\tilde{\lambda}_n$ obtained from the raw phonon
field configurations. There is a single dominant relative
variance. (b) Projection of the raw phonon field configurations onto the
plane of the two leading principal components.  The color bar indicates
the inverse temperature $\beta$ in units of $t$.  For small $\beta$
(high $T$), the pairs evolve with the usual topologies:  from a single
grouping centered at $(0,0)$ at small $\beta$ (high $T$) to
a pair of groupings at
larger $\beta$ (low $T$). 
(c) The quantified first leading component as a function
of $\beta$.  The dashed line indicates the QCP (separating
the two topologies) $\beta_c \approx 5.5$ which is close to values
obtained by conventional approaches. (d) The weight vector corresponding
to the first leading component, which shows a clear $(\pi,\pi)$ pattern.
\label{fig:Holsteineigenvalues} } 
\end{figure}

We used PCA to analyze the Holstein model at half filling on a 10$\times$10 square
lattice.  The PCA matrix was constructed from the phonon fields
$\{x_{i,\tau}\}$, for a single
fixed imaginary time slice $\tau$, providing $l=1000$ independent
configurations for each temperature.  Figure
\ref{fig:Holsteineigenvalues}\,(a) displays the relative variances, which
exhibit a single dominant component, suggesting the
existence of a dominant phonon displacement (or a charge) pattern.  The
projection of the first two principal components
[Fig.~\ref{fig:Holsteineigenvalues}\,(b)] has data points
(blue symbols) centered at the origin and high $T$, 
and split into two different clusters (red symbols) at low $T$.  This
splitting provides evidence of a phase
transition for a critical $T_{c}$.

\begin{figure}[t]
\includegraphics[width=0.98\columnwidth]{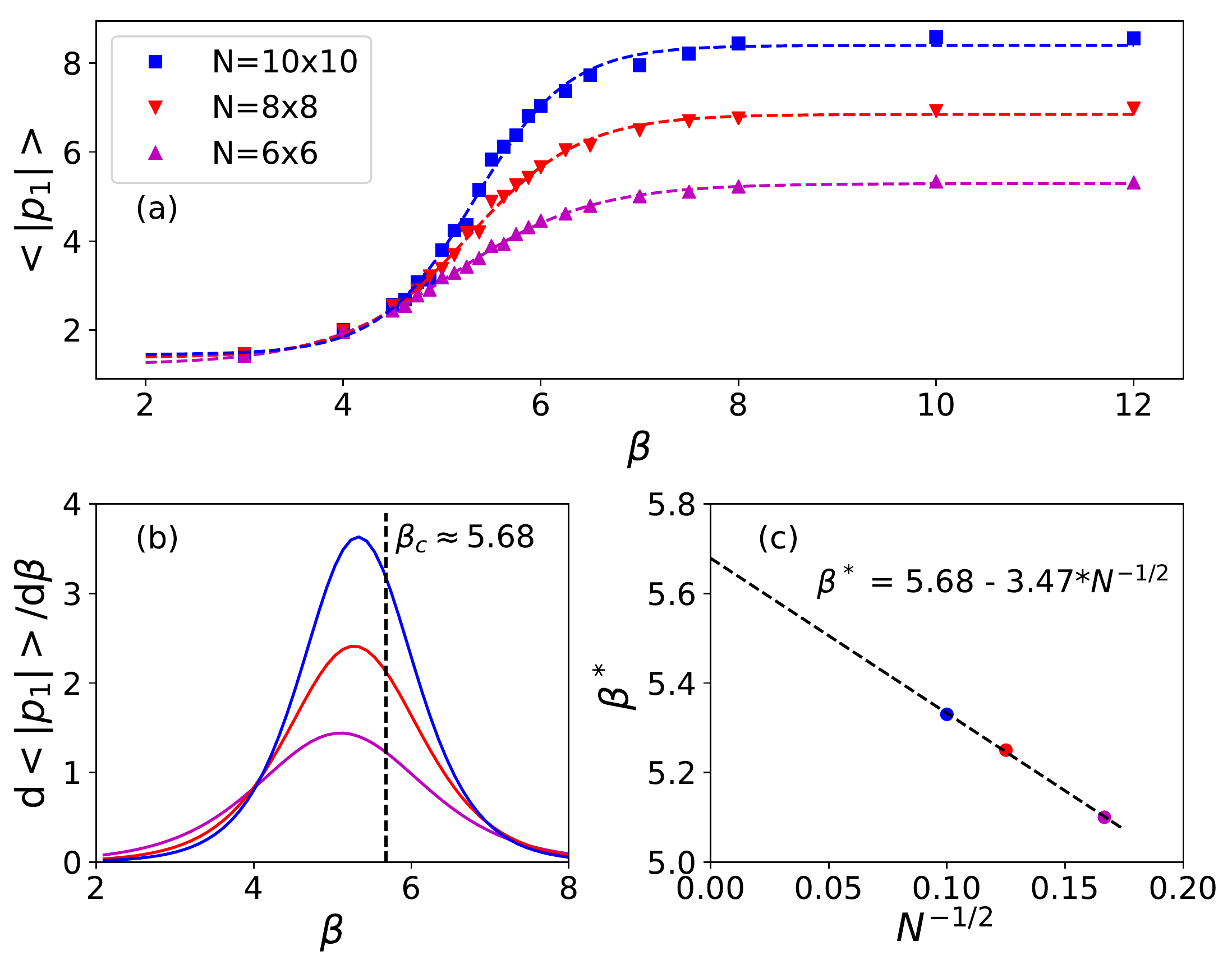} 
\caption{
Finite size scaling analysis of the Holstein model for $\lambda_{D}=\omega_{0}=1$. 
(a) Symbols are raw data for the quantified first leading component 
as a function of $\beta$, for different
lattice sizes. The dashed lines are guides to the eyes.
(b) Derivatives with respect to $\beta$ of numerical fits to the
raw data.
The dashed line marks the extrapolated critical value 
$\beta_c \approx 5.68$
obtained in panel (c) from a linear least-squares fit to the
peaks $\beta^*$ on different lattice sizes.
We assume the finite size correction goes as the inverse of the
linear system size.
\label{fig:Holsteinffs} } 
\end{figure}

\color{black}

In addition to the changes in the scatter plots of
Fig.~\ref{fig:Holsteineigenvalues}\,(b) with temperature, 
we also analyse the quantified
first leading component.  Figure \ref{fig:Holsteineigenvalues}\,(c) displays
the behavior of $\langle | p_{1} | \rangle$ as a function of $\beta$.
A sharp increase is evident for inverse temperature
in the range $ 4.5 \lesssim \beta \lesssim 6.5 $.  
Taking the midpoint of this range suggests
a PCA estimation for the critical temperature is $\beta_{c} \approx
5.5$.  Although previous DQMC results provide evidence of
$\beta_{c}=8$
(see Refs.\,\onlinecite{vekic92} and \onlinecite{noack91}, respectively),
a recent, more
accurate, analysis determined $\beta_{c}=6.0$ (Ref.\,\onlinecite{costa17b}),
in agreement with our PCA results.

To explore finite size effects, we compare simulations on
lattice sizes $N=6\times 6, \,\, 8\times 8$ and $10\times 10$.
In Fig.~\ref{fig:Holsteinffs}\,(a), the symbols are the PCA/DQMC results
for the quantified first 
leading component as a function of $\beta$,
while the dashed lines are their corresponding numerical fits.
Here we determine the inflection points by differentiating the
fitted curves, following Section\,\ref{Section:Honeycomb}.
Figure~\ref{fig:Holsteinffs}\,(b) displays the derivative of fitted curves,
with the inflection points (and ultimately, the ``critical temperatures")
being determined by their peaks.
In Fig.~\ref{fig:Holsteinffs}\,(c), we extract the critical 
value $\beta_c \approx 5.68$ using a linear least-squares fit on 
these inflection points. 
We have verified that fitting the data of
Fig.~\ref{fig:Holsteinffs}\,(a)
to different functional forms does not affect the values of 
$\beta^*$.

\vskip0.10in
\section{Conclusions}

This paper has extended prior work\cite{hu17} in which an unsupervised
learning approach based on the PCA was applied to a variety of {\it
classical} models of magnetism, to itinerant quantum Hamiltonians in two
dimensions.  The magnetic phase transitions in the Hubbard model on a
honeycomb lattice, the periodic Anderson model, and the one sixth filled
Lieb lattice, can all be observed via the evolution of the principal
components, even though the transitions are tuned in quite different
ways; via interaction strength, hybridization, and density respectively.

The similarities extend to the finite temperature Kosterlitz-Thouless
transition in the 2D attractive Hubbard Hamiltonian, which, like its
classical XY counterpart, proves less amenable to analysis.  In
contrast, the finite temperature CDW transition in the half-filled 2D
Holstein is well captured by the PCA, presumably because the broken
symmetry is discrete.

The similarities between PCA for classical and quantum models
of magnetism may be somewhat surprising, since, unlike 
short range spin models, the effective
classical degrees of freedom in DQMC are coupled by complicated, long
range interactions (the fermion determinants).  
Thus, in some ways, the application of PCA to
configurations provided by DQMC reported here, and in
Refs.\,\onlinecite{ching16,broecker16}, are similar to some of the original themes
explored in the interplay of learning and statistical mechanics which
considered (possibly frustrated) long range models.  

To see the full rotational symmetry of the magnetically
ordered ground states of the Hubbard model (e.g. the PAM) 
in the principal component distributions,
fermion spin configurations might be necessary to be fed into PCA directly.
As noted above, the particular form of the HS transformation will likely be relevant to the answer.
Complete resolution 
of this issue would be facilitated by a comparison
of results for different forms of the HS transformation, for example
by using the rotation symmetric form
introduced by Chen and Tremblay\cite{chen92},
but is beyond the scope of the present paper.

It remains to assess the extent of the advantages offered by machine
learning approaches to these transitions.  In the cases we have studied,
traditional approaches based on analysis of the known order parameters
would likely give a more accurate determination of the critical points.
These more precise values are usually achieved only following
a finite size scaling analysis,
as discussed in the last paragraph of the Holstein model section.
Finally, one should acknowledge that the more established methods have
been improved and refined over three decades.  Machine
learning techniques for DQMC 
await a similar development and improvement process.

In particular, an intriguing opportunity is offered in cases where the
sign problem makes the usual evaluation of a response function $\chi$
excessively noisy.  Because the PCA does not involve the computation of
the ratio $\langle \, \chi S \, \rangle/\langle \, S \, \rangle$, but
rather only the generation and analysis of configurations of the
Hubbard-Stratonovich field (with the absolute value of the determinants
as the weight), it seems possible that insight into transitions beyond
the sign problem might prove possible.  One knows that if the sign is
ignored, then the response functions can give incorrect information
about the physics (in the case of the 2D repulisive Hubbard model a
$d$-wave pairing amplitude which decreases as $T$ is lowered instead of
increasing)\cite{white89,loh05}.  Understanding whether a machine
learning analysis of the full space-time HS field configuration
generated with the absolute value of the sign is similarly misleading is
an open question.


\begin{acknowledgements}
This work was supported by the Department of Energy under grant number
DE-SC0014671 and by the Brazilian agencies Faperj and CNPq.
\end{acknowledgements}

\bibliography{Costa_learningQCP}

\end{document}